\documentclass[aps,prl,preprint,showpacs]{revtex4}
\usepackage{graphicx}
\usepackage{helvet}
\usepackage{amsmath,amssymb}
\usepackage{bm}

\begin{document}

\title{Insulating nature of strongly correlated massless Dirac fermions in an organic crystal}

\author{Dong Liu$^{1}$, Kyohei Ishikawa$^{1}$, Ryosuke Takehara$^{1}$, Kazuya Miyagawa$^{1}$, 
Masafumi Tamura$^{2}$, Kazushi Kanoda$^{1}$}
	
\affiliation{
$^{1}$ Department of Applied Physics, University of Tokyo, Bunkyo-ku, Tokyo, 113-8656, Japan \\
$^{2}$ Department of Physics, Faculty of Science and Technology, Tokyo University of Science, 
Noda, Chiba, 278-8510, Japan \\}

\date{\today}

\begin{abstract}
	Through resistivity measurements of an organic crystal hosting massless Dirac fermions with 
a charge-ordering instability, we reveal the effect of interactions among Dirac fermions on 
the charge transport. 
	A low-temperature resistivity upturn appears robustly irrespectively of pressure and 
is enhanced while approaching the critical pressure of charge ordering, indicating that 
the insulating behavior originates from short-range Coulomb interactions. 
	Observation of apparently vanishing gap in the charge-ordered phase accords with 
the theoretical prediction of the non-topological edge states.
\end{abstract}



\maketitle

	Strongly interacting electrons and massless Dirac fermions (DFs) are both of keen interest in 
condensed matter physics. 
	The interplay between interactions and massless nature is a prime issue in the physics of 
DFs and has been intensively studied for the single-layer carbon atoms, graphene \cite{Ref1, Ref2}. 
	Even before the successful separation of graphene, renormalization-group studies predicted that 
the Fermi velocity is logarithmically enhanced on approaching the Dirac point due to 
the unscreened long-range Coulomb repulsion among electrons \cite{Ref3}. 
	Actually, the quantum oscillations in suspended graphene proved the reshaping of 
the Dirac cone that results from the velocity renormalization \cite{Ref4}. 
	However, how the resistivity of interacting DFs behaves is under debate. 
	In the absence of interactions, the Dirac point conductance is theoretically given by 
the universal quantum value of $4e^{2}/\pi h$ \cite{Ref5} (with some modification in 
tilted cones \cite{Ref6}), which had been one of experimental targets in the early stage of 
the research \cite{Ref7}. 
	In the presence of interactions, it was predicted that the quasi-particle scattering rate of 
interacting DFs is proportional to temperature \cite{Ref8}, being distinct from 
the conventional behavior of Fermi liquids, and the Dirac-point conductance in 
clean graphene is suggested to increase logarithmically on cooling from 
the order of $4e^{2}/\pi h$ around room temperature \cite{Ref9}. 
	The experimental results, however, are not in line with the prediction; 
$\sigma$ at the Dirac point decreases at low temperatures, in many cases, following power laws of 
$T$ but with unsettled exponents \cite{Ref10, Ref11}. 
	These behaviors have been argued in the light of inhomogeneity or ripples in 
real graphene samples suspended or attached on 
substrates \cite{Ref10, Ref12, Ref13, Ref14, Ref15}, or broken valley symmetry under 
top-gating on the BN substrate \cite{Ref11}. 
	The environments appear influential in the low-energy charge transport in graphene. 
	It remains to be answered how the Dirac-point conductance is impacted by the correlation effect.

\begin{figure}
	\includegraphics[width=8.6cm]{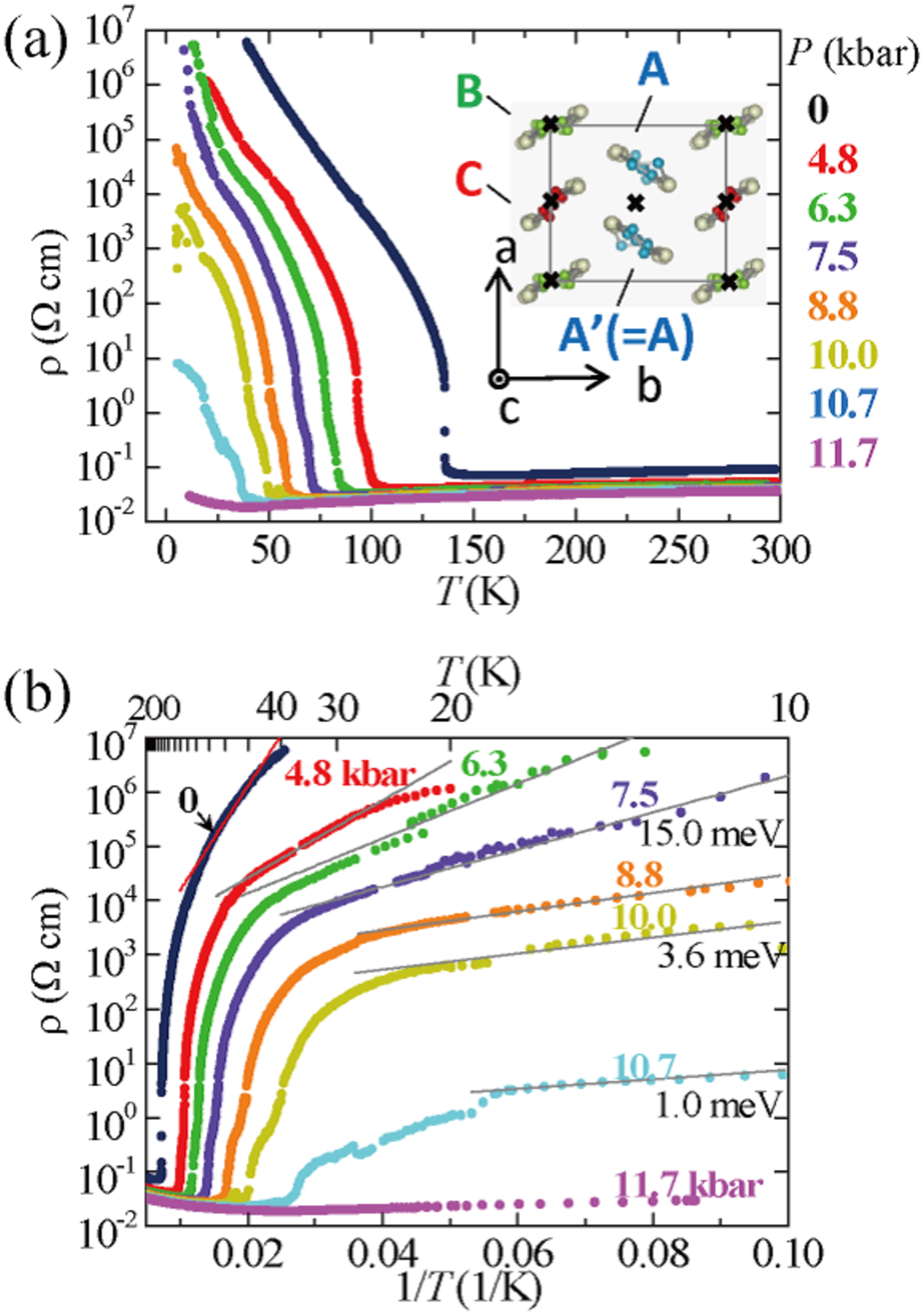}
	\caption{\label{Fig1} (Color online) Temperature-dependence of resistivity 
(a) and the corresponding Arrhenius plot 
(b) for $\alpha$-(BEDT-TTF)$_{2}$I$_{3}$ under different pressures. 
Inset of (a) shows the crystal structure of $\alpha$-(BEDT-TTF)$_{2}$I$_{3}$ viewed from $c$ axis, 
namely, the $a$-$b$ in-plane structure, in the conducting state. 
Crystallographically inequivalent molecular sites A(A'), B and C comprise a unit cell as 
indicated by the square. 
Black crosses indicate the position of inversion centers.}
\end{figure}

	We tackle this problem with another DF system, a quasi-two-dimensional organic conductor, 
$\alpha$-(BEDT-TTF)$_{2}$I$_{3}$ \cite{Ref16, Ref17} (abbreviated as $\alpha$-I$_{3}$ hereafter), 
which is composed of the conducting layers of (BEDT-TTF)$^{+1/2}$ cations hosting DFs 
(as shown in the inset of Fig.~\ref{Fig1}(a)) and the insulating layers of 
triiodide anions (I$_{3}$)$^{-1}$, where BEDT-TTF stands for bisethylenedithio-tetrathiafulvalene. 
	This is the first material that exhibits a charge-ordered (CO) state and 
a DF state right beside each other in the pressure-temperature phase diagram. 
	At ambient pressure, $\alpha$-I$_{3}$ undergoes a transition at 135 K from a conducting state to 
a CO state with the inversion symmetry broken, where charge is disproportionated between the two A sites in a unit cell as shown in the inset of Fig.~\ref{Fig1}(a) \cite{Ref18, Ref19, Ref20, Ref21, Ref39, Ref40}; however, when the CO is suppressed by 
pressure, a massless DF phase with the nature of 
strong correlation emerges \cite{Ref22, Ref23, Ref24}. 
	Compared with graphene, $\alpha$-I$_{3}$ is distinctive in the following respects: 
(i) A layered bulk crystal of $\alpha$-I$_{3}$ is free from both of 
structural deformation inevitable in free suspension and the influences of substrates, 
and thus the Dirac point could be approached in a super-clean condition; 
(ii) Owing to a fixed band filling (3/4) in the crystal, the Fermi energy is exactly located at 
the charge-neutral Dirac point; 
(iii) The strength of electron interactions is systematically varied near 
the CO-DF phase boundary by pressure; 
(iv) The Dirac cone in $\alpha$-I$_{3}$ is so largely tilted that 
the $k$-dependent Fermi velocity varies over approximately a ten-fold range \cite{Ref24}.

	Preceding transport studies of $\alpha$-I$_{3}$ revealed that 
the resistivity shows an anomalous upturn with a logarithmic increase on cooling \cite{Ref25}. 
	As the upturn is suppressed by carrier doping \cite{Ref26}, 
which makes the Fermi level shift from the Dirac point, 
the low-temperature upturn is characteristic of the Dirac point. 
	In the present work, we have explored the charge transport in $\alpha$-I$_{3}$ in 
a wide pressure range covering the CO and DF phases with particular focus on 
the critical region of the CO-DF transition to clarify the relationship between 
the transport anomaly and electron correlations.

	Single crystals of $\alpha$-I$_{3}$ were synthesized by 
the conventional electrochemical oxidization method. 
	The typical size of crystals is 1 mm $\times$ 0.3 mm in the $a$-$b$ conducting plane, 
and 10 to 100 $\mu$m in thickness (along $c$ axis). 
	The in-plane electrical resistivity was measured by the four-terminal method. 
	For applying hydrostatic pressures, we used two clamp-type pressure cells, 
a BeCu single-wall cell and a BeCu/NiCrAl dual-structured cell, for pressures up to 
20 kbar and 40 kbar, respectively. 
	As pressure-transmitting media, we used Daphne 7373 and 7474 oils, 
which keep hydrostaticity up to 20 kbar and 40 kbar, respectively \cite{Ref27}. 
	All the pressure values given below are the values at room temperature before cooling. 
	At ambient pressure, we performed measurements for 13 crystals, 
four of which were further used for experiments up to 13 kbar and one of 
the four was measured under increasing pressures up to 20 kbar. 
	We used two separate crystals for higher pressures of 25 to 40 kbar.

\begin{figure}
	\includegraphics[width=8.6cm]{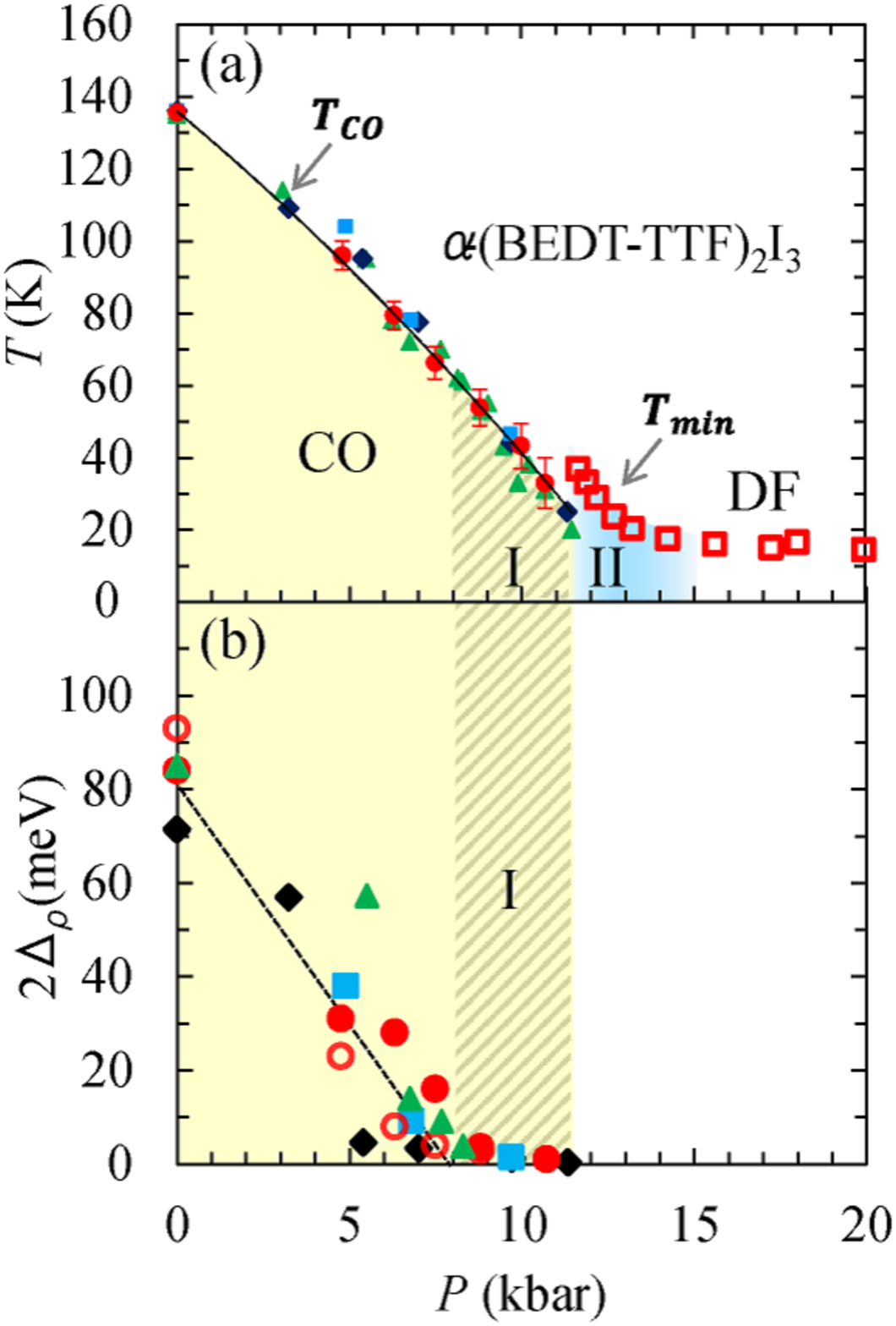}
	\caption{\label{Fig2} (Color online) (a) Pressure-temperature phase diagram. 
Different symbols indicate the charge-ordering temperatures $T_{\textrm{CO}}$ determined from 
different crystals. 
For the definition of $T_{\textrm{CO}}$, see text. 
The solid line is guide to the eye. 
$T_{\textrm{min}}$ indicates the temperature at which the resistivity takes a minimum in 
the Dirac fermion state under pressures above 11 kbar. 
$T_{\textrm{CO}}$ (red circles) and $T_{\textrm{min}}$ are from the same sample. 
Regions I and II are discussed in the text. 
(b) Pressure dependence of the activation energy of the Arrhenius plot in Fig. 1 (b). 
Filled and open red circles are from different surface places of the same crystal. 
The dashed line is guide to the eye.}
\end{figure}

	Figure~\ref{Fig1}(a) shows the temperature dependence of 
electrical resistivity $\rho$ under increasing pressures for one crystal; 
the Arrhenius plot of the $\rho$ is displayed in Fig.~\ref{Fig1}(b). 
	The overall characteristic is reproduced by all other crystals. 
	For $P <$ 11 kbar, the resistivity shows a steep increase at the CO transition. 
	The transition temperature, $T_{\textrm{CO}}$, is determined by 
a temperature giving a peak in the $-d(\textrm{Ln}\rho )/dT$ versus $T^{-1}$ curve and 
is shown against pressure in Fig.~\ref{Fig2}(a) together with 
the results of all other crystals studied up to 20 kbar. 
	The $T_{\textrm{CO}}$ decreases continuously with increasing pressure and 
eventually vanishes above 11 kbar. 
	Below $T_{\textrm{CO}}$, the temperature dependence of $\rho$ is not of 
the ideal Arrhenius type and somewhat sample-dependent, 
as reported earlier (ambient-pressure data) \cite{Ref28}; 
nevertheless, the activation energies of the Arrhenius plots, $\Delta_{\rho}$, 
for all the samples studied exhibit a systematic variation with pressure, 
as shown in Fig.~\ref{Fig2}(b). 
	The $\Delta_{\rho}$ decreases with increasing pressure more rapidly than 
$T_{\textrm{CO}}$ does and vanishes around 7-8 kbar. 
	Unexpectedly, the pressure dependence of $\Delta_{\rho}$ does not scale with 
that of $T_{\textrm{CO}}$ at all and there is a pressure range, 
as indicated by Region I in Fig.~\ref{Fig2}, where $\Delta_{\rho}$ vanishes 
while the charge ordering is identified by clear resistive anomalies at $T_{\textrm{CO}}$. 
	At approximately 11 kbar, $T_{\textrm{CO}}$ drops from 20 K to 0 K discontinuously, 
indicating a CO-DF phase transition of the first order. 
	These are consistent with the Raman study, 
which showed that the charge disproportionation ratio, 0.2:0.8 between the A and A' sites (Fig.~\ref{Fig1}(a) inset), at ambient pressure decreases at 
a rate of ca. 0.01/kbar with increasing the pressure up to 11 kbar and vanishes, at least, 
at 15 kbar \cite{Ref21}.

\begin{figure}
	\includegraphics[width=8.6cm]{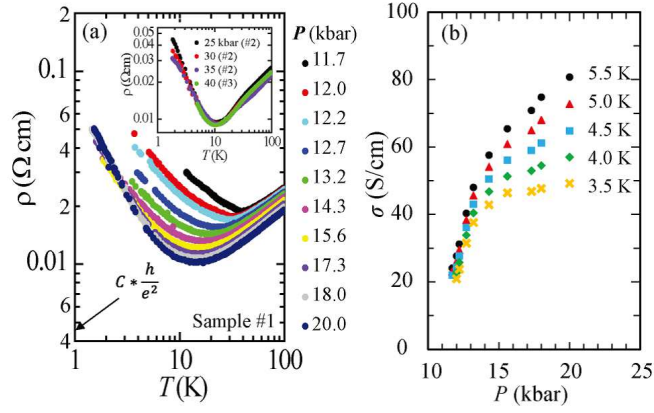}
	\caption{\label{Fig3} (Color online) (a) Temperature-dependence of resistivity in 
the Dirac fermion state above 11 kbar. 
Sample $\sharp$1 covers pressures up to 20 kbar (main panel). 
Sample $\sharp$.2 was measured at 25, 30, and 35 kbar, and sample $\sharp$3 was used for 
the measurements at 40 kbar (inset). 
The bulk resistivity corresponding to the quantum sheet resistance, C $\times$ ($h$/e$^{2}$)=
4.5 $\times$ 10$^{-3}$ Ohm cm, is indicated, where C = 1.7 nm is the lattice constant along 
the $c$-axis \cite{Ref38}] under pressure of around 20 kbar. 
(b) Conductivity versus pressure for fixed temperatures above 11 kbar 
in the low-temperature region, where resistivity upturn appears.}
\end{figure}

	For pressures above 11 kbar, the resistivity is characterized by metallic but 
weak temperature-dependences followed by anomalous upturns at low temperatures. 
	The high-temperature behavior, which reproduces the previous results \cite{Ref22}, 
is understood in terms of the compensation of the contrasting temperature dependences of 
the carrier density and mobility \cite{Ref22}. 
	We particularly focus on the low-temperature upturn observed in the whole pressure range for 
all the samples measured. 
	At high pressures above 15 kbar, the upturn is nearly pressure-insensitive, which 
is confirmed to persist up to 40 kbar, as shown in Fig.~\ref{Fig3}(a). 
	It is remarkable that below 15 kbar the low-temperature upturn is enhanced and 
the enhancement is in a near-critical manner near the CO-DF transition pressure, indicating the weak first-order nature of the transition at low temperatures, although 
$\rho$ at high temperatures is pressure-insensitive. 
	Concomitantly, the temperature at which resistivity takes a minimum, $T_{\textrm{min}}$, 
shows an accelerated increase as pressure is decreased from 15 kbar, 
as shown in Fig.~\ref{Fig2}(a). 
	The DF state is stable and pressure-insensitive at high pressures but changes its nature 
when the CO transition is approached. 
We denote this transient region as Region II in the phase diagram hereafter.

	First, we discuss the charge transport in the CO state. 
	At ambient pressure, the $\Delta_{\rho}$ values of the samples studied are in a rage of 
40$\pm$5 meV and thus 2$\Delta_{\rho}$ approximately accords with the optical gap, 
75 meV \cite{Ref29}. 
	However, the charge gap under pressure is highly unusual in 
that $\Delta_{\rho}$ is not scaled to $T_{\textrm{CO}}$ and in particular vanishes in 
Region I in spite of the resistivity anomalies signifying the bulk nature of the charge ordering. 
	The contradicting behaviors of $T_{\textrm{CO}}$ and $\Delta_{\rho}$ suggest 
extraordinarily low-energy charge excitations or the presence of 
tiny conducting portions or paths in the bulk CO background. 
	In connection with the former possibility, 
the $b$-aixs optical conductivity \cite{Ref29} and dielectric response \cite{Ref30} at 
ambient pressure exhibit low-energy excitations, which are discussed in terms of 
phason-like and domain-wall-like excitations \cite{Ref30}. 
	These excitations, however, are not pertinent to the dc conductivity, as indicated in 
the present results at ambient pressure and earlier \cite{Ref30}. 
	As the latter case, the coexistence of tiny DF domains due to the first-order transition or 
inhomogeneity in internal pressure is conceivable. 
	However, it is questionable that the DF state stabilized as a bulk phase 
above 11 kbar appear near ambient pressure even if it is a tiny volume. 
	Alternatively, Omori \textit{et al}. theoretically suggested the appearance of edge states in 
the CO phase \cite{Ref31}. 
	It is well known that DF materials are accompanied by peculiar edge/surface states. 
	Zigzag-type edge states in graphene with broken bulk inversion symmetry, acquiring massive nature, is suggested to vary from a gapped flat-band to gapless edge modes with valley-polarization by applying potential on the sample edge \cite{Ref32} or those with valley- and spin- polarization by turning on the edge ferromagnetism \cite{Ref41}. Edge transport is robust against impurities with smooth potentials because of the valley-polarization of the gapless edge states \cite{Ref41}.
	In $\alpha$-I$_{3}$, the CO state in the vicinity of the CO/DF phase boundary was also 
predicted to be characterized by massive Dirac fermions \cite{Ref33}. 
	The calculations of the edge states in the CO phase with 
incorporating the electron correlation found that the energy gap in the edge states formed in 
between upper and lower bands in the bulk decreases with the magnitude of Coulomb interactions, 
which is varied by pressure in experiments, and closes before the CD-DF phase boundary is reached. 
	This can contribute to the transport \cite{Ref31}, being fully consistent with 
the present observation (see Fig.~\ref{Fig2} (b)). 
	The predicted edge states are not topologically protected, being distinctive from 
those of topological insulators \cite{Ref34} and the spin-polarized zero-mode-Landau level of 
$\alpha$-I$_{3}$ \cite{Ref35}, and thus are sensitive to the roughness of the edge surfaces. 
	This explains why the absolute value of $\rho$ remains large. 
	A large drop in $\rho$ at 10.7 kbar is likely due to the coexistence of the DF phase just in 
the vicinity of the critical pressure.

	On entering into the DF state across the critical pressure of 11 kbar, 
the charge transport in the DF state is featured by 
the characteristic temperature $T_{\textrm{min}}$ and the magnitude of 
the associated resistivity upturn at $T < T_{\textrm{min}}$, both of 
which are profoundly enhanced near the critical pressure of the CO-DF transition. 
	The quantum sheet resistance, $h$/e$^{2}$ = 25.8 k Ohm, corresponds to the bulk resistivity, 
4.5 $\times 10^{-3}$ Ohm cm, in the present case. 
	The resistivity appears to decrease toward this value before turning to 
the increase (see Fig.~\ref{Fig3}(a)). 
	The pressure evolution of conductivity at $T < T_{\textrm{min}}$ is such that 
it decreases toward vanishing at 10.5-11 kbar, which is close to the CO-DF critical pressure, 
as seen in Fig.~\ref{Fig3}(b). 
	The pressure dependence of the upturn in region II indicates that 
the electron correlation is responsible for the upturn because it is enhanced in 
a quasi-critical manner near the correlation-induced transition to the CO. 
	In the framework of the renormalization of unscreened long-range Coulomb interactions based on 
the Weyl equation, the conductivity is predicted to increase logarithmically against 
temperature decrease \cite{Ref9}, in sharp contrast to the experimental features. 
	A disorder-induced localization, a conceivable origin of the resistivity increase, 
is unlikely to explain the clear pressure-dependence of the upturn and the absence of 
the negative magneto-resistance \cite{Ref26} as observed in graphene \cite{Ref10}.

	Thus, the present results strongly indicate that the resistivity upturn at low-$T$ is due to 
interactions among DFs and invoke a notion beyond the conventional framework to 
treat the interaction effects on DFs. 
	It is known that in real $\alpha$-I$_{3}$ crystals the I$_{3}^{-1}$ deficiency of 
several ppm causes chemical potential to deviate from the Dirac point by the order of 
a few Kelvin, corresponding to submillivolt gating \cite{Ref25}; 
nevertheless, the upturn was sample-independent, indicating that a few Kelvin around 
the Dirac point is a characteristic energy scale where the picture of 
the renormalized conical dispersion may break down; 
something unexpected beyond the renormalization may happen in the low energies, 
the scale of which can be more specifically dictated by $T_{\textrm{min}}$ of 10 K. 
	The low-temperature resistivity increase may be an indication of a pseudogap formation in 
charge channel, which is enhanced while approaching the DF/CO phase transition at 
$P_{\textrm{C}}$ $\sim$ 11 kbar before forming a real gap in the CO phase. 
	The CO is stabilized by the intersite Coulomb interactions, which is not incorporated in 
modeling interacting DFs in terms of the Weyl equation, which assumes that DFs travel in 
a continuum. In real systems, however, they are on lattices, of which the transfer integrals sets 
the upper limit of kinetic energy, being exceeded by the short-range interactions in the CO state. 
	The significance of the short-range interactions even in 
the DF regime is theoretically argued in terms of the Hubbard model \cite{Ref36}. 
	As the renormalization effect of the long-range Coulomb interactions on the electron velocity is expected 
to persist without an anomaly at a particular pressure, the short-range part of 
the Coulomb interactions is likely a key to the anomalous resistivity increase, 
leading to a possible consequence that the ground state of interacting DFs on 
lattices is insulating.

	In conclusion, we investigated the low-temperature resistivity of interacting Dirac fermions in 
an organic crystal, $\alpha$-(BEDT-TTF)$_{2}$I$_{3}$, which is free from 
the complexities associated with free suspension and attached substrates, 
around the critical pressure of a transition to the charge-ordered phase. 
	In the Dirac fermion phase, increases in resistivity at low temperatures are confirmed as 
a robust feature of Dirac fermions and found to be enhanced in a near-critical manner 
while approaching the charge-ordering transition. 
	This is an indication that the resistivity upturn toward an insulating ground state 
is caused by the interactions of Dirac fermions --- a novel feature of the interacting Dirac liquid 
situated in a quantum critical point \cite{Ref37}, which shows up 
when the environmental influences are minimized. 
	In the charge-ordered phase, an apparent suppression of the charge gap with 
pressure increase is revealed and found most reasonably explainable in terms of the emergence of 
the edge states suggested theoretically.

	We thank M. Hirata and M. Koshino for useful discussions. 
	This work was supported in part by JSPS KAKENHI under Grant Nos. 20110002, 25220709, 24654101.

\newpage

\newpage
\newpage

\newpage

\end{document}